\documentclass[sigconf]{acmart}
\AtBeginDocument{%
  }
\usepackage{amsmath} 
\usepackage{multirow}
\acmISBN{978-1-4503-XXXX-X/2018/06}

\copyrightyear{2025}
\acmYear{2025}
\setcopyright{acmlicensed}\acmConference[SCID '25]{The 2nd Workshop on Security-Centric Strategies for Combating Information Disorder}{August 25--29, 2025}{Hanoi, Vietnam}
\acmBooktitle{The 2nd Workshop on Security-Centric Strategies for Combating Information Disorder (SCID '25), August 25--29, 2025, Hanoi, Vietnam}
\acmDOI{10.1145/3709020.3734834}
\acmISBN{979-8-4007-1417-7/2025/08}

\hypersetup{
  colorlinks=true,
  linkcolor=blue,
  citecolor=blue,
  urlcolor=blue,
  pdfborder={0 0 0} 
}

\begin{document}

\title{E-FreeM$^2$: \underline{E}fficient Training-\underline{Free} \underline{M}ulti-Scale and Cross-Modal News Verification via \underline{M}LLMs}


\author{Van-Hoang Phan}
\authornote{These authors contributed equally to this research.}
\orcid{0009-0003-8615-435X}
\affiliation{
  \institution{University of Science}
  \city{Ho Chi Minh City}
  \country{Vietnam}
}
\affiliation{
  \institution{Vietnam National University}
  \city{Ho Chi Minh City}
  \country{Vietnam}
}
\email{21120459@student.hcmus.edu.vn}

\author{Long-Khanh Pham}
\authornotemark[1]
\orcid{0009-0004-6197-3428}
\affiliation{
  \institution{University of Science}
  \city{Ho Chi Minh City}
  \country{Vietnam}
}
\affiliation{
  \institution{Vietnam National University}
  \city{Ho Chi Minh City}
  \country{Vietnam}
}
\affiliation{%
  \institution{FPT Software AI Center}
  \country{Vietnam}
}
\email{21120479@student.hcmus.edu.vn}

\author{Dang Vu}
\authornotemark[1]
\orcid{0009-0003-3268-024X}
\affiliation{%
  \institution{University of Science}
  \city{Ho Chi Minh City}
  \country{Vietnam}
}
\affiliation{
  \institution{Vietnam National University}
  \city{Ho Chi Minh City}
  \country{Vietnam}
}
\affiliation{%
  \institution{FPT Software AI Center}
  \country{Vietnam}
}
\email{23C15023@student.hcmus.edu.vn}
\email{dangvqm@fpt.com}

\author{Anh-Duy Tran}

\orcid{0000-0002-8036-954X}
\affiliation{%
 \institution{DistriNet, KU Leuven}
 \city{Leuven}
 \country{Belgium}
}
\email{anh-duy.tran@kuleuven.be}


\author{Minh-Son Dao}
\orcid{0000-0003-3044-8175}
\affiliation{%
  \institution{National Institute of Information and}
  \institution{ Communications Technology}
  \city{Tokyo}
  \country{Japan}
}
\email{dao@nict.go.jp}


\begin{abstract}
The rapid spread of misinformation in mobile and wireless networks presents critical security challenges. This study introduces a training-free, retrieval-based multimodal fact verification system that leverages pretrained vision-language models and large language models for credibility assessment. By dynamically retrieving and cross-referencing trusted data sources, our approach mitigates vulnerabilities of traditional training-based models, such as adversarial attacks and data poisoning. Additionally, its lightweight design enables seamless edge device integration without extensive on-device processing. Experiments on two fact-checking benchmarks achieve SOTA results,  confirming its effectiveness in misinformation detection and its robustness against various attack vectors, highlighting its potential to enhance security in mobile and wireless communication environments.
\end{abstract}


\maketitle

\section{Introduction}


In recent years, the proliferation of media manipulation technologies, such as Deepfakes, has captured significant attention due to their qualities and their role in amplifying the spread of false information \cite{shao2023detecting, tolosana2020deepfakes, wang2022deepfake, zhao2021multi}. These technologies, powered by advancements in artificial intelligence, including video, image, and speech synthesis, have raised concerns about their potential misuse in generating fake news and facilitating financial fraud, prompting responses from both regulatory authorities and industry stakeholders to mitigate their impact. However, one of News Verification’s most pervasive yet understudied challenges is out-of-context (OOC) misinformation, where authentic, unaltered images are repurposed with misleading or false narratives \cite{fazio2020out}. For instance, during the recent Israel-Hamas conflict, social media platforms witnessed numerous cases of OOC misinformation, with old images from unrelated conflicts or even video game footage falsely presented as current events \cite{deepfakes-example}. Unlike manipulated media, OOC misinformation is challenging to detect because the visual content remains authentic, and the deception arises solely from the misleading context provided by accompanying text.

Addressing this ``infodemic'' requires multifaceted strategies, with news verification playing a pivotal role through fact-checking and debunking false claims \cite{nikolov2020teamalexclefcheckthat, covid19}. Misinformation, which is shared unintentionally, and disinformation, deliberately disseminated to deceive often for political motives, both thrive in digital ecosystems  \cite{Coronavirus}. Social media platforms, such as TikTok, have emerged as significant vectors for disinformation, particularly among younger audiences, through formats like Deepfakes and less sophisticated ``Cheapfakes'' \cite{tiktok}. As these platforms amplify the reach of false narratives, developing robust news verification techniques becomes essential to ensure the integrity of information. This research aims to address the critical challenges of detecting and mitigating OOC misinformation and manipulated media, leveraging advancements in artificial intelligence to enhance the accuracy and reliability of news verification systems.

Recent advancements in Out-of-Context (OOC) detection—which focuses on determining whether an authentic image is accurately represented within a given claim—have increasingly relied on external information retrieval through custom search engines such as the Google API \cite{papadopoulos2023red, yuan2023support, zhang2023ecenet}. These methods have demonstrated significant improvements by incorporating additional contextual evidence. In contrast, earlier approaches that lacked external data integration \cite{aneja2023cosmos, mu2023self, 10.1145/3652583.3657597, 10.1145/3652583.3657596} depended solely on the intrinsic features of image-claim pairs, which limited their effectiveness in verifying claims.
Professional fact-checkers typically assess both internal image features and external contextual information to determine the validity of an image-claim relationship. However, leveraging retrieved evidence efficiently remains a major challenge in OOC detection, particularly in identifying subtle inconsistencies between images and associated claims. To address this, recent studies \cite{abdelnabi2022open, qi2024sniffer, tahmasebi2024multimodal, 10.1145/3652583.3657597} have investigated the integration of retrieved evidence into classifiers, pretrained models, and large vision-language models (LVLMs). While these enhancements improve accuracy, they often come at the cost of increased architectural complexity and computational overhead.


Several prior studies \cite{10.1145/3652583.3657597, abdelnabi2022open, tahmasebi2024multimodal, tonglet-etal-2024-image} have made significant contributions to fact verification in multimodal tasks. By leveraging pretrained knowledge, reasoning capabilities, and generative abilities, these approaches effectively detect factual inconsistencies in image-text pairs while generating coherent, natural language-based explanations.
Building on this foundation, recent works \cite{qi2024sniffer} enhance fact-checking models by augmenting input data with external information retrieved via tool-based methods before applying instruction tuning for training OOC misinformation detectors. However, training models on curated fact-checking datasets for OOC misinformation detection presents several challenges. One major drawback of relying on fixed datasets is their inability to keep up with evolving misinformation trends. Since these datasets must undergo collection, preprocessing, and publication before they can be used, they quickly become outdated. Additionally, they are vulnerable to static data poisoning attacks, reducing their adaptability, security, and effectiveness in real-time fact-checking.
Another critical challenge is the high computational cost of training or adapting Multimodal Large Language Models (MLLMs). Furthermore, to remain effective against rapidly evolving misinformation, frequent retraining is necessary, further increasing computational and storage overhead.

To address these challenges, we introduce E-FreeM$^2$, a specialized Multimodal Large Language Model (MLLM) designed for real-time OOC misinformation detection with high accuracy. Our approach enhances verification by leveraging a cross-modal, multi-scale framework to retrieve and generate high-quality candidate information.
E-FreeM$^2$ first conducts external searches to gather additional evidence, employing cross-modal strategies for candidate ranking and multi-stage filtering to refine the retrieved data. The final decision-making process follows a two-stage chain-of-thought reasoning mechanism: the first stage validates the retrieved candidates against the input, while the second stage detects OOC misinformation based on consolidated judgments and explanations from the previous step.


Our contribution can be summarized as follows:
\begin{itemize}
    \item  \textbf{Cross-Modal Data Pipeline}: We design a novel cross-modal data pipeline that operates on both local and global scales, assisted by a Filter Module, to transform OOC image-text pairs into high-quality candidate sets. These refined candidates serve as the foundation for accurate judgments and coherent explanations.
    \item \textbf{Training-Free Adaptation for MLLMs}: We propose E-FreeM$^2$, a practical, training-free approach to adapt existing MLLMs for out-of-context misinformation detection using a two-stage chain-of-thought instruction mechanism. Our method efficiently leverages retrieved cross-modal candidates, effectively modeling both internal image-text relationships and external claim-evidence connections, enabling simultaneous OOC detection and explanation.
    \item \textbf{SOTA Performance Without Training Data}: Comprehensive experiments demonstrate that E-FreeM$^2$ significantly outperforms state-of-the-art (SOTA) methods in detection accuracy, all while requiring no training data or additional computational resources. Moreover, it generates precise and persuasive explanations, enhancing the interpretability of its decisions.
\end{itemize}

The remainder of this paper is structured as follows. Section 2 reviews related work on Out-of-Context detection and Chain-of-Thought prompting in multimodal models. Subsequently, Section 3 introduces the proposed E-FreeM$^2$ framework, outlining its evidence retrieval pipeline and two-stage reasoning process. Section 4 then describes the experimental setup, including datasets, metrics, results, and ablation studies. Finally, Sections 5 and 6 discuss limitations, suggest future directions, and summarize the main contributions.

\section{Related work}
\subsection{Out-of-Context Detection}
With the rapid surge of multimodal misinformation across social networks, researchers have increasingly focused on developing solutions for multimodal fact-checking, particularly in detecting out-of-context (OOC) misinformation. Some methods \cite{gu2024learning, luo2021newsclippings} utilize knowledge-rich pre-trained models to conduct internal checking for the given image-text pair. While these methods have demonstrated strong performance in OOC detection, they often overlook the integration of external knowledge relevant to the image-claim pair. Consequently, such models struggle to effectively capture logical or factual inconsistencies, leading to limitations in their learning.

Alternative strategies incorporate external resources for verification. For example, prior studies \cite{jaiswal2017multimedia, jaiswal2019aird, sabir2018deep} utilize reference datasets containing unmodified claims to approximate real-world knowledge, facilitating OOC detection by comparing the given claim with retrieved references. Similarly, \cite{abdelnabi2022open} retrieves web-based evidence for both textual and visual inputs separately and then evaluates their consistency with the claim across multiple modalities. Furthermore, Qi et al. \cite{qi2024sniffer} were the first to integrate the LVLM model with Vicuna-13B to address OOC misinformation within the realm of generative AI. In contrast, our approach emphasizes efficiency by adopting a training-free framework alongside multi-scale cross-modal candidate retrieval.
\begin{figure*}
    \centering
    \includegraphics[width=0.8\linewidth]{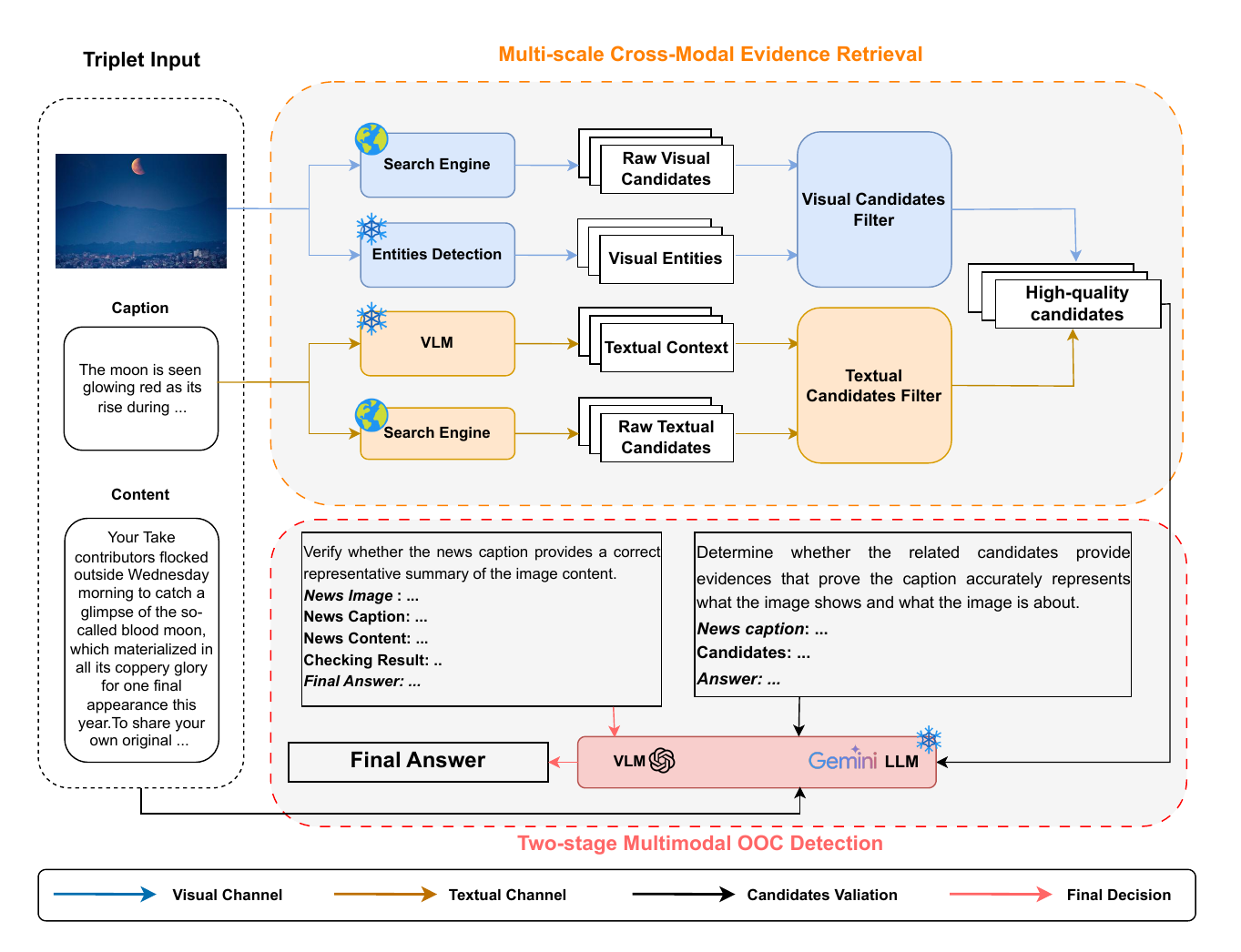}
    \caption{\textbf{Overview of E-FreeM$^2$ Framework.} Our efficient training-free multi-scale and cross-modal news verification system operates through two main components: (1) \textbf{Multi-Scale Cross-Modal Evidence Retrieval} - employing dual textual and visual retrieval pipelines with hierarchical similarity-based filtering and visual-centric ranking to obtain contextually relevant evidence from external sources; and (2) \textbf{Two-Stage Multimodal OOC Detection} - Stage 1 performs evidence-guided multimodal verification using structured reasoning to assess claim-evidence alignment, while Stage 2 conducts final multimodal decision reasoning through comprehensive synthesis of external evidence and direct visual analysis.}
    \label{fig:OverallMethod}
\end{figure*}
\subsection{Chain-of-thought in MLLMs}
Multimodal Large Language Models (MLLMs) have been developed to align perception with language models, allowing for cross-modal transfer of knowledge between language and multimodal domains \cite{huang2023language}. The introduction of datasets like SPIQA for multimodal question answering and evaluation strategies like Chain-of-Thought (CoT) have further improved model performance in interpreting complex information within scientific papers \cite{pramanick2025spiqadatasetmultimodalquestion}. Additionally, studies have explored the use of CoT prompting systems for Image Quality Assessment (IQA) in MLLMs, demonstrating the effectiveness of in-context retrieval and fine-grained assessment \cite{wu2024comprehensivestudymultimodallarge}. Furthermore, the integration of Chain of Thought methods in multimodal reasoning tasks has shown promising results in various domains, such as sports understanding and green building design decision-making \cite{xia2024sportucomprehensivesportsunderstanding, li2024question}.

\section{Efficient Training-Free Multi-Scale and Cross-Modal News Verification via MLLMs: Design and Implementation}

\subsection{Multi-Scale Cross-Modal Evidence Retrieval}  
To effectively detect out-of-context (OOC) misinformation, we propose a \textbf{Multi-Scale Cross-Modal Evidence Retrieval} framework, which consists of three key components: \textbf{Cross-Modal Retrieval}, \textbf{Similarity-Based Evidence Refinement}, and \textbf{Visual-Centric Ranking}. This framework systematically retrieves, filters, and ranks external evidence from both textual and visual modalities. Unlike traditional unimodal retrieval strategies that rely solely on either text-based or image-based matching, our approach integrates both modalities to capture inter-modal inconsistencies more effectively. This ensures that retrieved evidence is contextually relevant, thereby improving both fact verification and explanatory reasoning.  
\paragraph{Cross-Modal Retrieval}  
Our system employs two complementary retrieval pipelines to acquire diverse and contextually relevant evidence.  
\begin{itemize}
    \item \textbf{Textual Retrieval Pipeline}: Given an input claim, the textual pipeline extracts relevant captions, article snippets, and metadata from external sources using claim-conditioned queries.
    \item \textbf{Visual Retrieval Pipeline}: To assess visual consistency, an image-based reverse search is performed to retrieve visually similar images along with their associated metadata, such as source domains, contextual captions, and publication timestamps.  
\end{itemize}
\paragraph{Multi-Stage Similarity-Based Evidence Refinement}  
To enhance retrieval accuracy, we introduce a modular \textbf{Filter Module} designed to perform hierarchical evidence refinement using both modality-specific and cross-modal similarity metrics. This two-step process ensures that only high-quality, relevant evidence is retained for downstream reasoning. The filtering consists of two main stages: \textbf{Modality-Specific Similarity Filtering} and \textbf{Contextual Relevance Filtering}.  

In the first stage, \textbf{Modality-Specific Similarity Filtering}, the system assesses the relevance of both textual and visual evidence by computing similarity scores.  
\begin{itemize}
    \item \textbf{Textual Similarity Filtering}: Semantic similarity is computed between the input claim and candidate textual evidence (e.g., captions, snippets) using transformer-based dense embeddings (all-MiniLM-L6-v2) and cosine similarity scoring.  
    \item 
\textbf{Visual Similarity Filtering}: Perceptual similarity is computed using Vision Transformer (ViT) embeddings, followed by cosine similarity matching against the query image.  

\end{itemize}
Candidates are retained if their similarity scores exceed a predefined threshold:  
\begin{equation}
    S_{\text{similarity}} \geq \theta, \quad \theta = 0.7
\end{equation}  
ensuring that only strongly relevant evidence is preserved for further refinement.  

The second stage, \textbf{Contextual Relevance Filtering}, applies additional criteria to eliminate irrelevant, unreliable, or redundant information. This process involves:  
\begin{itemize}
    \item \textbf{Domain Filtering}: Retaining evidence only from verified and reputable sources (e.g., trusted news outlets) to ensure reliability.  
    \item \textbf{Language Filtering}: Excluding non-English evidence to maintain linguistic consistency in downstream reasoning, preventing misinterpretation due to translation errors.  
    \item \textbf{Redundancy Reduction}: Removing duplicate or near-duplicate entries through a domain-title deduplication heuristic to prevent bias from redundant sources.  
\end{itemize}
\paragraph{Visual-Centric Ranking}  
While both textual and visual similarities are computed, the ranking mechanism prioritizes \textbf{visual similarity} as the dominant factor in final candidate selection:  
\begin{equation}
    S_{\text{final}} = S_{\text{visual}}
\end{equation}  

This prioritization is based on the observation that visually similar images tend to preserve real-world contextual integrity and are less susceptible to manipulative textual reframing. Although semantic similarity remains a secondary criterion, it is used to resolve borderline cases and refine rankings within visually similar candidate sets. Following the filtering and ranking stages, the refined evidence set is consolidated into a unified multimodal representation that includes both text-based and image-based evidence. This structured set is then processed by the reasoning module, which utilizes it for OOC detection and explanatory output generation.  


\subsection{Two-stage Multimodal OOC Detection}
To enable robust and efficient Out-of-Context (OOC) misinformation detection, we propose a Two-Stage Multimodal OOC Detection framework that integrates external evidence retrieval with structured reasoning mechanisms. This approach is inspired by professional fact-checking workflows, where retrieved knowledge is combined with scene-level image understanding to assess the veracity of claims. By leveraging cross-modal retrieval and stepwise multimodal reasoning, the system ensures accurate, interpretable, and contextually aware decision-making.

\paragraph{Stage 1: Evidence-Guided Multimodal Verification.}  
In the first stage, we focus on verifying the consistency between the claim, expressed through the news caption, and external candidate evidence retrieved from trusted sources. A Multimodal Large Language Model (Gemini) is utilized to assess the alignment between the textual content of the caption and the retrieved candidate information. Each candidate is represented using a structured schema that incorporates semantic features and similarity metrics, capturing how well the external evidence supports or challenges the caption’s narrative. The model performs a hierarchical verification process that integrates similarity analysis with contextual reasoning. Rather than relying on isolated features, it jointly considers image-caption consistency, the credibility of external sources, and the degree to which the retrieved evidence substantiates or challenges the claim. Particular attention is paid to verifying critical elements such as depicted entities, temporal and spatial references, and the overall coherence between the news image, caption, and external evidence.


\paragraph{Stage 2: Final Multimodal Decision Reasoning.}

In the second stage, the system performs a comprehensive synthesis of multimodal insights to reach a final determination regarding the out-of-context status of the claim. A dedicated decision module, powered by a Multimodal Large Language Model (GPT-4o mini), integrates the verification outcome from Stage 1 with direct visual reasoning over the news image. The model combines external evidence results with scene-level image understanding to perform structured reasoning. It extracts key semantic elements from the image, such as entities, locations, objects, and contextual cues, and cross-references these against the claims and implications present in the caption. Additionally, the model assesses whether the caption introduces misleading narratives or selective framing that may distort the intended message conveyed by the image.


\paragraph{Specifically, we outline the CoT techniques integrated into the final checking prompts:}

\begin{itemize}
    \item \textbf{Step-by-step Analysis:} The prompts require users to follow a series of logical steps, breaking down the task into manageable parts, such as caption matching, evidence verification, and contextual alignment \cite{wei2023chainofthoughtpromptingelicitsreasoning}. This decomposition of the task enhances clarity and supports structured reasoning.
    \item \textbf{Contextual Reasoning:} The prompts emphasize the importance of understanding the broader context surrounding the image \cite{qiao2023reasoninglanguagemodelprompting}. This includes verifying details such as the location, time, people, and specific events depicted in the image. Contextual reasoning ensures that captions are not only checked for direct matches but are also evaluated against the broader situational factors.
    \item \textbf{Verification of Evidence:} A key CoT technique used in the prompts is the validation of evidence from multiple sources. The instructions encourage the use of supporting evidence (e.g., textual descriptions, image similarity scores, domain authority) to substantiate claims made by the caption. This evidence-based reasoning enhances the accuracy of the final decision.
    \item \textbf{Confidence Scoring}: The prompts incorporate a confidence scoring system (ranging from 0 to 10) to reflect the certainty of the decision \cite{portillo-wightman-etal-2023-strength}. CoT techniques are evident in this step, as each piece of evidence (such as the similarity of the caption or image) contributes incrementally to the overall confidence score, thus facilitating a more nuanced decision-making process and avoiding hallucination.
\end{itemize}

Overall, the incorporation of chain-of-thought methodologies in multimodal language models has shown significant potential in enhancing model performance and addressing complex reasoning tasks across various domains.


\section{Experiment}
\subsection{Experimental Setup}
\subsubsection{Dataset}
We evaluate our framework on the NewsCLIPpings dataset, the largest benchmark for out-of-context (OOC) misinformation detection to date. NewsCLIPpings is constructed from VisualNews, a large-scale corpus comprising image-caption pairs sourced from four major news outlets: The Guardian, BBC, USA Today, and The Washington Post. The OOC instances are generated by replacing original images with visually or semantically similar images retrieved from unrelated news events, creating challenging mismatches between images and captions.
Following prior work, we evaluate exclusively on the Merged/Balance test set (7,264 samples), which ensures a balanced mix of in-context and out-of-context cases across retrieval strategies. Using a training-free approach, we report overall accuracy, along with separate scores for Out-of-Context (OOC) and Not-Out-of-Context (NOOC) cases.
\subsubsection{Evaluation Metrics}

\textbf{Accuracy:} measures the overall correctness of the model's predictions by evaluating the proportion of correctly classified samples. It is reported across three categories: (1) \textbf{All}, which represents the overall accuracy across all samples, (2) \textbf{OOC (Out-of-Context)}, which measures accuracy for  detecting fake news, and (3) \textbf{NOOC (Not Out-of-Context)}, which evaluates accuracy on identifying real news.

\subsection{Performance Comparison}
\subsubsection{OOC Detection Comparison}
\begin{table}[h]
    \centering
    \begin{tabular}{lccc}
        \toprule
        \textbf{Method} & \textbf{All} & \textbf{OOC} & \textbf{NOOC} \\
        \midrule
        SAFE  \cite{zhou2020similarity}              & 52.8  & 54.8  & 52.0  \\
        EANN      \cite{wang2018eann}          & 58.1  & 61.8  & 56.2  \\
        VisualBERT  \cite{li2019visualbert}        & 58.6  & 38.9  & 78.4  \\
        CLIP \cite{radford2021learning}              & 66.0  & 64.3  & 67.7  \\
        DT-Transformer \cite{abdelnabi2022open}      & 77.1  & 74.8  & 75.6  \\
        CCN      \cite{papadopoulos2023synthetic}           & 84.7  & 84.8  & 84.5  \\
        Neu-Sym detector \cite{zhang2024interpretabledetectionoutofcontextmisinformation}    & 68.2  & -     & -     \\
        \midrule
        SNIFFER             & 88.4  &  86.9 & \textbf{91.8}  \\
        E-FreeM$^2$ (Ours)  & \textbf{90.0}  & \textbf{90.67}  & 89.49 \\
        \bottomrule
    \end{tabular}
    \caption{Accuracy (\%) comparison of different methods.}
    \label{tab:accuracy_comparison}
\end{table}

Table \ref{tab:accuracy_comparison} shows that E-FreeM$^2$ achieves the highest overall accuracy (90.0\%), outperforming all baseline methods, including SNIFFER (88.8\%), CCN (84.7\%), and DT-Transformer (77.1\%). Notably, E-FreeM$^2$ excels in Out-of-Context (OOC) detection with 90.67\%, surpassing SNIFFER (86.9\%) by a significant margin, while maintaining competitive performance in Not-Out-of-Context (NOOC) cases (89.49\%). These results demonstrate that E-FreeM$^2$ effectively distinguishes between in-context and out-of-context samples, validating its robustness across different retrieval strategies.

\subsubsection{Efficiency Comparison}

Table \ref{tab:trainable_params} highlights the efficiency of our method compared to existing approaches in terms of trainable parameters and inference time. Unlike VisualBERT (110M), CLIP (149M), and SNIFFER (99M), which require a substantial number of trainable parameters, our method operates with \textbf{zero trainable parameters}, demonstrating its lightweight and training-free nature. Additionally, our approach achieves an inference time of \textbf{12.77 seconds} when classifying a sample, emphasizing its computational efficiency, as shown in Table \ref{tab:accuracy_comparison}.
\begin{table}[h]
    \centering
    \begin{tabular}{lc}
        \toprule
        \textbf{Method} & \textbf{Trainable Params} \\
        \midrule
        VisualBERT \cite{li2019visualbert} & 110M \\
        CLIP \cite{radford2021learning} & 149M \\
        SNIFFER & 99M \\
        \midrule
        Ours & \textbf{0M} \\
        \bottomrule
    \end{tabular}

    \caption{Trainable parameters and inference time of different methods.}
    \label{tab:trainable_params}
\end{table}

\subsection{Ablation Studies}
\subsubsection{Ranking and Filter Strategies}

\begin{figure}[H]
    \centering
    \includegraphics[width=0.8\linewidth]{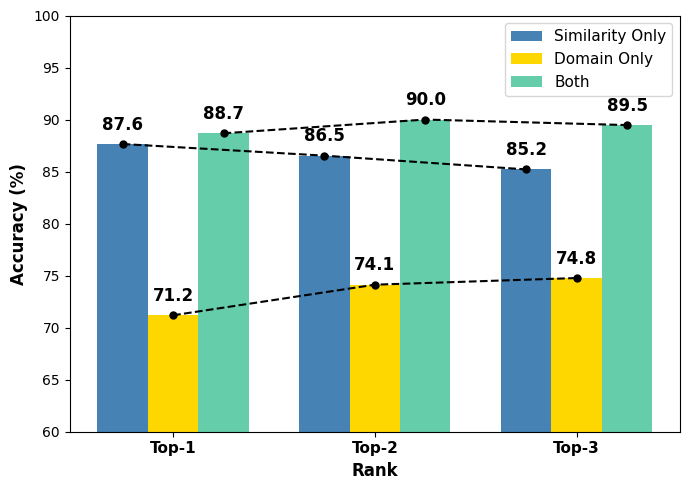}
    \caption{Ablation results with filter strategies}
    \label{fig:filterranking}
\end{figure}

To evaluate the effectiveness of different evidence filtering strategies, we conducted a performance comparison using three ranking approaches: Similarity Only, Domain Only, and Both (combining similarity and domain filtering). The results were visualized in Figure \ref{fig:filterranking}, where the x-axis represents ranking positions (Top-1, Top-2, and Top-3), and the y-axis represents accuracy (\%) in detecting out-of-context (OOC) misinformation. The results indicate that combining similarity-based and domain-based filtering yields the highest accuracy across all ranking positions, outperforming approaches that rely solely on either method. While similarity-based ranking provides strong contextual relevance, integrating domain reliability further enhances performance, demonstrating the importance of a hybrid filtering strategy for effective OOC misinformation detection.
\subsubsection{Training Dataset  Utilization}
\begin{figure}[H]
    \centering
    \includegraphics[width=0.8\linewidth]{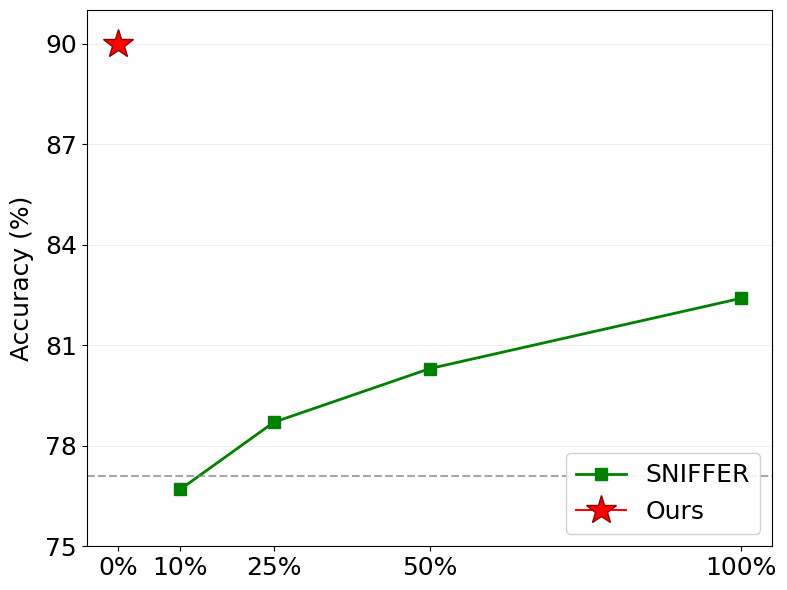}
    \caption{Percentage of used training set of NewsCLIPpings\cite{luo2021newsclippings}}
    \label{fig:data_augmentation}
\end{figure}

To compare the effectiveness of our approach against SNIFFER, Figure \ref{fig:data_augmentation} plots accuracy (\%) on the y-axis and the proportion of training data used on the x-axis, where SNIFFER's performance improves with more data, while our method (marked with a red star) achieves higher accuracy without requiring any training data. 
The results demonstrate that our approach significantly outperforms SNIFFER, highlighting its efficacy as a training-free solution for OOC misinformation detection.

\subsubsection{Effects of filtering methods}
\begin{table}[H]
    \centering
    \begin{tabular}{lccc}
    \hline
    Method & Accuracy \\
    \hline
    E-FreeM$^2$ (Ours)     & \textbf{90.0}  \\
    - w/o Image Evidences & 76.48 \\
    - w/o Text Evidences  & 77.05 \\
    - w/o Domain Filters         & 56.46 \\
\hline
\end{tabular}
\caption{Evaluation results for different evidence filtering methods.}
\label{tab:evidence_ablation}
\end{table}
\vspace{-10mm}
Table \ref{tab:evidence_ablation} shows the impact of evidence filtering on model performance. The full model achieves the highest accuracy (90.0\%). Removing either image evidence (accuracy drops to 76.48\%) or text evidence (accuracy drops to 77.05\%) significantly degrades performance, indicating both are crucial. Conversely, low-quality evidence severely harms performance: removing domain filters, which eliminate such evidence, causes accuracy to plummet to 56.46\%. This underscores the critical role of domain filtering in maintaining contextual understanding and high accuracy by removing detrimental inputs.

\section{Limitation and Discussion}
While E-FreeM² demonstrates strong performance, certain limitations should be acknowledged for future improvement. Firstly, the framework's reliance on external search engines for evidence retrieval introduces a dependency on the availability and quality of these external sources. Secondly, the current implementation is primarily focused on English-language content, potentially limiting its applicability in other linguistic contexts.  Thirdly, while effective against OOC misinformation, its robustness against other sophisticated manipulations, such as deepfakes, has not been extensively evaluated. Furthermore, the visual-centric ranking, while beneficial, might underperform in scenarios where textual cues are more critical than visual similarity.  Finally, the efficiency of the multi-stage filtering process could be further optimized to reduce latency in real-time applications.

\section{Conclusion}
This paper presents a novel framework for training-free, cross-modal fact verification. Out-of-context misinformation detection is the task of identifying misleading image-caption pairs that misrepresent visual content. Our approach, E-FreeM², utilizes multi-scale retrieval pipelines and a dual-stage reasoning mechanism based on large multimodal language models to analyze contextual consistency between images, captions, and externally retrieved evidence. By combining cross-modal evidence refinement and structured chain-of-thought reasoning, our system delivers highly accurate and interpretable verification results.

Future directions include extending  E-FreeM²'s capabilities to handle multilingual content and exploring defenses against more sophisticated manipulations such as deepfakes. We also aim to further optimize latency and retrieval mechanisms to enable even faster fact-checking on edge devices. By harnessing the full potential of multimodal reasoning and retrieval, we hope to build even more adaptable, secure, and robust solutions to combat misinformation in dynamic media environments.

\begin{acks} 
\indent
We would like to thank the Vingroup Innovation Foundation (VINIF) for partially funding this research under the grant number \\
VINIF.2021.JM01.N2.

\end{acks}
\bibliographystyle{ACM-Reference-Format}
\bibliography{ref}

\end{document}